%% file: narita.tex
\documentclass[12pt]{article}
\usepackage[]{amsmath}
\def\dfrac#1#2{{\displaystyle\frac{#1}{#2}}}

\def\cfrac#1#2{\dfrac{\mathstrut #1}{#2}}
\begin{document}
\baselineskip 20pt
\centerline{\Large{\bf On the Behaviour of Solutions to}}
\centerline{\Large{\bf Discrete Time Lotka-Volterra Equation}}

\centerline{\Large{-- Integrable Case --}}
\vglue 1cm
\centerline{Yasuaki NARITA \footnote{E-mail: ynarita@phys.metro-u.ac.jp},\quad Satoru SAITO \footnote{E-mail: saito@phys.metro-u.ac.jp}}

\centerline{Noriko SAITOH \footnote{E-mail: saitoh@lam.osu.sci.ynu.ac.jp} and\quad Katsuhiko YOSHIDA\footnote{E-mail: yoshida@phys.metro-u.ac.jp}}
\vglue 0.5cm
\begin{center}{\it 
${}^{1,2}$Department of Physics, Tokyo Metropolitan University\\
1-1 Minamiohsawa, Hachiohji, Tokyo, Japan 192-0397}
\end{center}

\begin{center}{\it 
${}^3$Department of Applied Mathematics, Yokohama National University\\
Hodogaya-ku, Yokohama, Japan 240-8501}
\end{center}

\begin{center}{\it 
${}^4$School of Science, Kitasato University\\
1-15-1 Kitasato, Sagamihara, Kanagawa, Japan 228-8555}
\vspace{5mm}
\end{center}

\begin{center}{\large \bf Abstract}\end{center}
The time evolution of a class of completely integrable discrete Lotka-Volterra system is shown not unique but have two different ways chosen randomly at every step of generation. This uncertainty is consistent with the existence of constants of motion and disappears in both continuous time and ultra discrete limits.

\section{Introduction}

Generally speaking we are not able to specify all of solutions of a given equation. A deterministic equation is either completely integrable or nonintegrable. In case of a non-integrable system, some of orbits converge into stable orbits, otherwise behave chaotic.  Moreover, in spite of the deterministic nature of equation, chaotic orbits are very sensitive on their initial conditions. Namely one can not predict behaviour of orbits which had infinitesimally small difference of initial values.

Soliton equations are, therefore, distinct among other nonlinear equations. They are completely integrable under wide class of initial data. Solutions are often given explicitly in terms of $\tau$ functions. One of important features of these systems is the existence of discrete analogue of equations which preserve integrability. This is remarkable since there are infinitly many possible ways of discretization which violate integrability and create chaos. We are interested in understanding criteria which distinguish integrable discrete systems from nonintegrable ones. In the case of 2nd order ordinary differential equations, Painlev\'e property plays one of such roles. 

In this paper we consider a discrete analogue of the Painlev\'e property and show explicitly how it works to characterize integrability. We study, in particular, $N$-point Lotka-Volterra map. It is equivalent to the discrete time Toda lattice, hence is completely integrable. Such integrable systems are usually believed that their orbits are also deterministic and insensitive on their initial values. Namely starting from an arbitrary initial value an orbit is determined uniquely.

The purpose of this paper is to show that above view of a completely integrable system is not always correct. By analysing the discrete Lotka-Volterra equation we will show that the orbit is not determined uniquely but there arise two ways of orbit to be chosen at every step of time evolution. This unexpected behaviour owes to the appearance of a square root function. Hence the situation is similar to the Painlev\'e property for ordinary differential equations, in which an appearance of a moving singularity implies nonintegrability of the equation. We can show explicitly the branch point of the square root function moves depending on initial values when the equation is perturbed by a small parameter. If the parameter vanishes, the square root is resolved as expected, but two valuedness remains and causes uncertainty in its evolution.

The equation we discuss in this paper is the following type of discrete time Lotka-Volterra (d-LV) equation :
\begin{eqnarray}
 x_n^{t+\delta}(1-\delta x_{n-1}^{t+\delta})
 &=& x_n^t(1-\delta x_{n+1}^t) ,
 \hspace{5mm} n=1,\cdots,N
 \label{eq:Nd-LV} \\
 x_0^t = x_N^t,~~x_1^t &=& x_{N+1}^t~~~\mbox{(periodic condition)} ,
 \label{eq:period1}
\end{eqnarray}
where $\delta$ is the minimum size of time step.

In the limit $\delta \rightarrow 0$ , 
(\ref{eq:Nd-LV}) becomes the continuous time Lotka-Volterra equation :
\begin{eqnarray}
 \frac{d}{dt}x_n(t) = x_n(t)[x_{n-1}(t)-x_{n+1}(t)] ,
 \label{eq:N-LV}
\end{eqnarray}
which is more familier and known integrable. 
Conversely equation (\ref{eq:Nd-LV}) is a natural discretization of (\ref{eq:N-LV}) in the sense that it is obtained from (\ref{eq:N-LV}) within the framework of Hirota bilinear formalism, which preserves integrability.

The complete integrability of the equation (\ref{eq:Nd-LV}) itself is guaranteed by the existence of sufficient number of conserved quantities. For our purpose it will be more useful to notice the connection of d-LV (\ref{eq:Nd-LV}) to the discrete time Toda lattice,
\begin{equation}
 \left\{ 
 \begin{array}{ccl}
 I_n(t+\delta)V_n(t+\delta) &=& I_{n+1}(t)V_n(t) , \\
 \\
 I_n(t)-I_n(t+\delta) &=& \delta^2[V_{n-1}(t+\delta)-V_n(t)] ,
 \end{array}
 \right.
 \label{eq:d-Toda}
\end{equation}
by the Miura transformation\cite{HT}
\begin{equation}
 \left\{
  \begin{array}{ccl}
  V_n(t) &=& x_{2n}^t x_{2n+1}^t , \\
  \\
  I_n(t) &=& (1-\delta x_{2n-1}^t)(1-\delta x_{2n}^t).
 \end{array}
 \right.
  \label{eq:Miura-T2}
\end{equation}

The discrete time Toda lattice is a completely integrable system which has been well studied. Many useful information can be translated from the Toda lattice to our system. At the same time all results we obtain for the d-LV in the following discussions must be true also for the discrete time Toda lattice.

Recently integrable discretization has been further extended to ultra discretization. Namely associated to variety of discrete soliton equations there exist integrable equations whose dependent variables are also discrete. Starting from certain class of discrete soliton equations we can derive so called ultra discrete equations in a systematic way. In this way the ultra discrete analogue of d-LV was discovered in \cite{TTMS} :
\begin{eqnarray}
 d^{t+1}_n - d^t_n
 = \max[0,d^{t}_{n+1}-1]-\max[0,d^{t+1}_{n-1}-1]~~,~~
 n = 1,2,\cdots,N .
\label{NUd-LV}
\end{eqnarray}
We will also discuss this equation from the same point of view.

We are going to examine in detail the behaviour of solutions of $N$ point discrete Lotka-Volterra system in two separete papers. In the present paper we concentrate our attention to integrable case. Although solutions have been given explicitly in terms of $\tau$ functions, we will show that an actual orbit is not deterministic. At every step of time evolution there are possibility to jump from one stable orbit into another. Two different orbits are characterized by the same constants of motion. This unexpected behaviour exists only in the case of discrete map. Therefore we will show how this uncertain behaviour ceases to occur in the continuous time evolution. We are also interested in whether both types of solutions remain in the ultra discrete limit. In conclusion we will find that the orbit is determined uniquely.

Our study of the discrete Lotka-Volterra system can be extended to nonintegrable case by introducing a small deformation parameter. We can see in detail behaviour of the singularities at every step of time evolution. In addition a Julia set will turn out to be a proper object to study global feature of singularities. We can see how the Julia set disappears from the complex plane of dependent variables in the integrable limit. We will discuss all these in our succeeding paper.

\section{3-point Lotka-Volterra system}

First we concentrate our argument to 3-point (N=3) case, but most of the results which we
discuss below can be extended to arbitary N cases. Since we are interested in studying difference of variables before and after one step of time evolution, we set $\delta=1$ and denote 
$$
(x,y,z):=(x_1^t,x_2^t,x_3^t),\qquad (x',y',z'):=(x_1^{t+1},x_2^{t+1},x_3^{t+1}).
$$
Under this simplification of notation the 3-point d-LV equation is 
\begin{subequations}
\begin{eqnarray}
x'(1-z') &=& x(1-y) ,\label{x'=X}\\ y'(1-x') &=& y(1-z) ,\label{y'=Y}\\ z'(1-y') &=& z(1-x) .
\label{eq:3-LV}
\end{eqnarray}
\end{subequations}
Applying the scale transformation $(x,y,z)\rightarrow
(\delta x,\delta y, \delta z)$, we can restore the minimum time lemgth $\delta$
and recover the equation (\ref{eq:Nd-LV}) for the $3$-point case. Hence it is sufficient to consider only $\delta=1$
case (7), unless the continuous limit is taken.
\subsection{Two types of map}

In order to see the time evolution of the system we solve (7) for $(x',y',z')$ as functions of $(x,y,z)$. 
For convenience we denote the right hand sides of (\ref{x'=X}), (\ref{y'=Y}) and (\ref{eq:3-LV}) by
\begin{eqnarray}
X=x(1-y),\quad Y=y(1-z),\quad Z=z(1-x) .
\label{eq:XYZ}
\end{eqnarray}
Then the successive substitutions of (\ref{y'=Y}) and (\ref{eq:3-LV}) into (\ref{x'=X}) yield
\begin{eqnarray}
 x' &=& \cfrac{X}{1-z'}
     =  \cfrac{X}{1-\cfrac{Z}{1-y'}}
     =  \cfrac{X}{1-\cfrac{Z}{1-\cfrac{Y}{1-x'}}} \\
    &=& X \cfrac{1-Y -x'}{1-Y-Z-(1-Z)x'} .
\end{eqnarray}
Solving this for $x'$ we obtain 
\begin{eqnarray}
 x' = \cfrac{1+X-Y-Z \pm \sqrt{(1-X-Y-Z)^2-4XYZ}}{2(1-Z)} ~.
 \label{eq:x'=pm}
\end{eqnarray}

From this elementary calculas it is apparent that the orbit of the map is not unique. The singularity of the map associated to the appearance of the square root function arises at every step of the map. Since there is no reason for a solution in (\ref{eq:x'=pm}) to choose one of two signs of the square root, the map is two valued, hence not unique, at every step of the time evolution. The uniqueness of the map is violated unless the argument in the square root vanishes by itself irrespect to the values of $x,\ y,\ z$. But our result (\ref{eq:x'=pm}) shows that it vanishes only if $x,\ y,\ z$ are in some relation. 

How this could be compatible with the complete integrability of the equation. In order to clarify this point, let us substitute (\ref{eq:XYZ}) to (\ref{eq:x'=pm}) and see what happens. First we notice that the square root is resolved, i.e.,
\begin{eqnarray}
  \sqrt{(1-X-Y-Z)^2-4XYZ}=|xyz-(1-x)(1-y)(1-z)|,
\end{eqnarray}  
hence the map is not singular. This property resembles the Painlev\'{e} property of ordinary differential equations. To make this point clear we must compare above result with the case where $X,\ Y,\ Z$ differ from (\ref{eq:XYZ}). We will, however, postpone this argument to our succeeding  paper.

If we substitute this result into (\ref{eq:x'=pm}), we obtain
\begin{eqnarray}
x'= \left\{ 
		 \begin{array}{c}
		  x \cfrac{1-y+yz}{1-z+zx} \\
		  \\
	      1-y .
	     \end{array}
	     \right.
\label{2 solutions}
\end{eqnarray}
Hence we are left with the problem of non-uniqueness of the map. Both in (\ref{2 solutions}) are candidates of solution to (7) with equal importance. At every step of evolution there is no reason to choose any particular one out of two solutions. If we repeat the map $m$ steps there will arise $2^m$ possible orbits which we can choose both with same weight. An orbit obtained by solving (7) seems to become non-deterministic.

Now we ask how this could be compatible with complete integrability of d-LV. This will be understood if we solve (7) also for $y'$ and $z'$ in similar way as we did for $x'$ case. Because d-LV equation has cyclic symmetry under the rotation of $(x,y,z)$,
the result is 
\begin{eqnarray}
 \left(\begin{array}{c} x'\\y'\\z' \end{array}\right)~
 &=& \left(\begin{array}{c} x\cfrac{1-y+yz}{1-z+zx} \\ {}\\
                            y\cfrac{1-z+zx}{1-x+xy} \\ {}\\
                            z\cfrac{1-x+xy}{1-y+yz}
 \end{array}\right) \hspace{5mm} A\mbox{-type}
 \hspace{5mm} , \hspace{5mm}
 \left(\begin{array}{c} 1-y\\1-z\\1-x \end{array}\right)
 \hspace{5mm} B\mbox{-type}
\label{AB}
\end{eqnarray}
Hereafter we call these maps A-type and B-type respectively.

We must notice that there are not eight but only two types of the map at every step of evolution. If the minus sign is chosen in (\ref{eq:x'=pm}), the same sign must be chosen also for $y'$ and $z'$ to obtain a consistent answer. Do we still get $2^m$ different maps after $m$ steps? This is not so because of the special character of the B-type solution. Namely it is a map of reflection. If we repeat the map six times a point goes back to the original place. Moreover two successive map of B after the A-type map is equivalent to the two successive map of B followed by A-type of map. We can write them symbolically as
$$
B^6=1 ,\qquad AB^2=B^2A.
$$
Similarly we can find many such relations. Therefore the number of possible orbits will be reduced to smaller, but not to one.

\subsection{Conserved quantities}

Now we consider relation between conserved quantities and the two types of the map. In the 3-point case, conserved quantities are easily found,
\begin{eqnarray}
 C &\equiv& X+Y+Z = x + y + z - xy - yz - zx \\
 D &\equiv& X Y Z = x y z (1-x)(1-y)(1-z)  \hspace{2mm} ,
\end{eqnarray}
which are obtained simply by summing or multiplying the left hand side of (7), respectively. There is no other quantity independent from these two and being constant of motion. These two constants restrict orbits of the motion to some curves in the three dimensional real space $(x,y,z)$. 

To see the orbits more explicitly it will be more convenient to introduce new constants $r,s$ defined from $C$ and $D$ by
\begin{eqnarray}
 r s &=& D \nonumber\\
 r + s &=& 1 - C.
\label{1-C}
\end{eqnarray}
Writing them explicitly we have
\begin{eqnarray}
 r \equiv xyz ,\qquad s \equiv (1-x)(1-y)(1-z).
\end{eqnarray}
The constraint of $r$ to $r=a\ (\mbox{const})$ forms a hyperboloid. Similarly $s=b\ (\mbox{const})$ forms another hyperboloid. Hence the orbits of the map are bound into intersections of these two hyperboloids. 

We now observe that if the initial condition $(r,s)=(a,b)$ fixes orbits for a set of constants $(C,D)$, (\ref{1-C}) shows that $(s,r)=(a,b)$ also fixes orbits with the same values of $(C,D)$. It amounts to exchange $(x,y,z)$ with $(1-x,1-y,1-z)$. They form different intersections which are point reflection of each other with the center of reflection at $(x,y,z)=(1/2, 1/2, 1/2)$. Therefore for a given set of constants $(C,D)$, there are two sets of separate orbits, which we will call $O_+$ and $O_-$. 
\begin{figure}[tb]
\begin{center}
 \input{hozon.tex}
\end{center}
 \caption{Example of orbit (time steps 300)}
 \label{fig:hozon}
\end{figure}
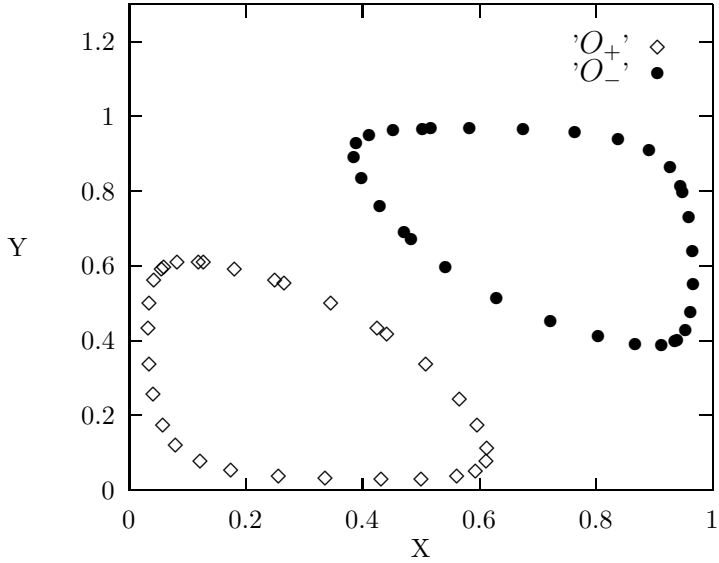
Figure \ref{fig:hozon} illustrates an example of these maps, 
which are projected on the $x,y$ plane. 
In this figure initial values are chosen at $(x,y,z)=(0.06,0.6,0.17)$ which draw $O_+$ with $(r,s)=(0.00612,0.31208)$. On the other hand $O_-$ is drawn with $(r,s)=(0.31208,0.00612)$ by starting from $(x,y,z)=(0.94,0.4,0.83)$.

We are now ready to see correspondence between the orbits drawn in the picture and our solutions in (\ref{AB}). By looking at the A-type solution, we find that $(r,s)$ remain the same after the map. On the other hand a B-type solution exchanges $(r,s)$ to $(s,r)$. Needless to say both maps are allowed since the constants of motion $C,D$ remain unchanged. Thus we obtain the following correspondence
\begin{itemize}
 \item $A-type \hspace{5mm}  r \rightarrow r ~~,~~ s \rightarrow s$
~~~(r,s are unchanged)
 \item $B-type \hspace{5mm}  r \rightarrow s ~~,~~ s \rightarrow r$
~~~(r,s are exchanged)
\end{itemize}

Starting from particular initial values of $(x,y,z)$, say (0.06,\ 0.6,\ 0.17), and if there is no B-type solution, we will find only $O_+$ orbit.

\subsection{Continuous time limit}

In section 2.1 we have seen d-LV eq. has two types of map. We are interested in their behaviour in the continuous time limit. The d-LV equation itself turns to the continuous time LV equation (\ref{eq:N-LV}) in the $\delta\rightarrow 0$ limit. This fact seems rather paradoxical in the first sight, because the uniqueness of solutions must be true in the case of ordinary differential equations. Namely there is only one orbit starting from a particular initial value. We like to see how the correspondence between discrete system and continuous system holds in our case. We will try to understand this problem from two different view. 

First we look at the solutions (\ref{AB}) in detail when $\delta$ is small. For this purpose we restore $\delta$ in (\ref{2 solutions}),
\begin{equation}
x^{t+\delta}=\left\{
\begin{array}{rl}
	x^t\cfrac{1-\delta y^t+\delta^2y^tz^t}{1-\delta z^t+\delta^2z^tx^t}\qquad &:{\rm A-type}\\
 \\
{1\over\delta}-y^t\qquad &:{\rm B-type}.
\end{array}
\right.
\label{A type with delta}
\end{equation}
Expanding the both sides of the A-type solution for small $\delta$ we find
$$
x^t+\delta{dx^t\over dt}+o(\delta^2)=x^t(1-\delta y^t+\delta z^t+o(\delta^2))
$$
Hence in the first order of $\delta$ we obtain the continuous time LV equation (\ref{eq:N-LV}). The same arguments are true for $y^t$ and $z^t$. This means that the A-type solutions satisfy the LV equation (3) in the continuous time limit. As for the B-type solution all points are mapped to infinity when $\delta$ turns to zero. In other words the B-type solution has a meaning only if $\delta$ is finite. In this sense the solution becomes unique in the continuous time limit. 

The second view to clarify the behaviour of solutions is to see the dependence on $\delta$ of the cross section of the hyperboloids specified by the constants of motion. We have seen that there are two different kinds of cross sections, which we called $O_\pm$. Under a variance of $\delta$ we will find that the orbits $O_-$ move far away to infinity as $\delta$ becomes small, whereas the other orbits $O_+$ remain in finite region.

\begin{figure}[tb]
\begin{center}
 \input{del.tex}
\end{center}
\caption{Change of orbits under variance of parameter $\delta$}
\label{fig:2}
\end{figure}
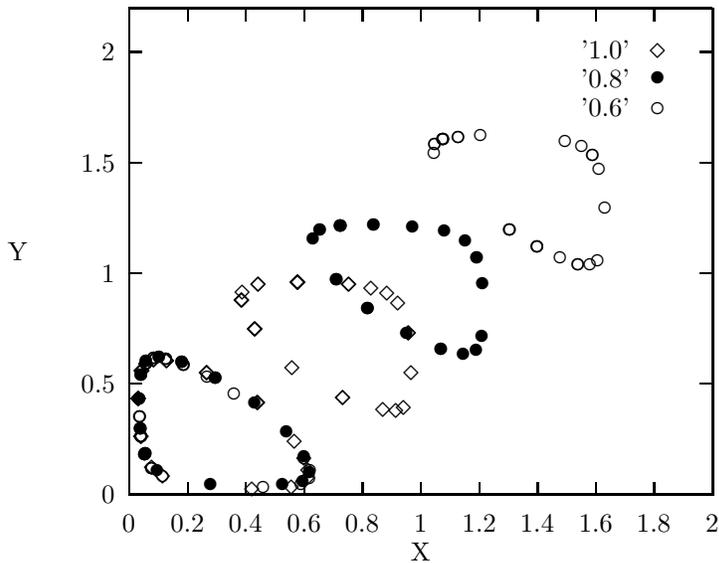

Figure \ref{fig:2} shows such an example of change of orbits for  $\delta=1.0,\ 0.8,\ 0.6$ 
respectively. Initial values are fixed at $(x,y,z)=(0.06,\ 0.6,\ 0.17)$.

\section{N-point Lotka-Volterra system}
We can generalize most of the results for 3-point discrete Lotka-Volterra system which we found in the previous section to arbitrary $N$ point case.
\subsection{Two types of map}
The equation which we concern in this section is (\ref{eq:Nd-LV}) with arbitrary $N$. Let us simplify, for a while, the notation by putting
\begin{equation}
x'_n=\delta x_n^{t+\delta},\qquad x_n=\delta x_n^t
\end{equation}
and the equation of motion
\begin{equation}
x'_n(1-x'_{n-1})=x_n(1-x_{n+1}),\qquad n=1,2,\cdots, N.
\label{eq:Nd-LV simple}
\end{equation}
We also denote the right hand side of (\ref{eq:Nd-LV simple}) by
\begin{equation}
X_n:=x_n(1-x_{n+1}),\qquad n=1,2,\cdots, N,
\label{X_n=}
\end{equation}
and solve (\ref{eq:Nd-LV simple}) for $x'_1$ in terms of $(X_1,X_2,\cdots ,X_N)$. Repeating the same procedure as in the case of 3-point d-LV we find
\begin{eqnarray}
 x'_1 &=& \cfrac{X_1}{1-\cfrac{X_{N}}{1-\cfrac{X_{N-1}}{~~~\cfrac{\ddots}{1-\cfrac{X_{2}}{1-x'_1}}} }} \\
 &=& X_1 \cfrac{c_{N-1}-a_{N-1}x'_1}{c_N-a_N x'_1}
 \label{eq:x1'}
\end{eqnarray}
where $a_k,\ c_k$ satisfy the recurrence relations
\begin{eqnarray}
 a_{k+1} &=& a_k -a_{k-1}X_{k+1}, \hspace{10mm} (a_0,a_1) = (0,1), \\
 c_{k+1} &=& c_k -c_{k-1}X_{k+1}, \hspace{11mm} (c_0,c_1) = (1,1),
\end{eqnarray}
or solving them explicitly
\begin{equation}
 a_k=a_{(3,k)},\qquad c_k=a_{(2,k)}
\end{equation}
where
\begin{equation}
a_{(i,k)}:=\sum_{l=0}^{l_{max}}C_l^{(i,k)},\qquad 
l_{max}=\left\{ \begin{array}{ll}
                     [(k-i+2)/2]  & k\ge i \\
                    {[(N-i+k+2)/2]} & k<i
                \end{array}\right.
\label{a_(i,k)}
\end{equation}
and 
\begin{equation}
C_0^{(i,k)}:=1,\qquad C_l^{(i,k)}:=(-1)^l\sum_{j_1,j_2,\cdots,j_l}^{(i,k)}X_{j_1} X_{j_2} \cdots X_{j_l},\quad l\ne 0.
\end{equation}
The summation $\sum_{j_1,j_2,\cdots,j_l}^{(i,k)}$ must be taken over $j_a$'s satisfying 
\begin{eqnarray}
j_1,j_2,\cdots, j_l\in
\left\{
\begin{array}{cc}
(i,i+1,\cdots,k),\qquad & i\le k\\ 
(i,i+1,\cdots, N,1,2,\cdots, k),\qquad & i\ge k
\end{array}
\right.
\end{eqnarray}
$$
 |j_a-j_b|\ge 2,
$$
and $[k]$ is the maximum integer which is equal or smaller than $k$. 
\\
From (\ref{eq:x1'})
$x'_1$ becomes 
\begin{eqnarray}
x'_1=\cfrac{b \pm\sqrt{b^2-4a_Nc_{N-1}X_1}}{2a_N},\qquad 
b:=c_N+a_{N-1}X_1.
\label{x'_1}
\end{eqnarray}
If we define $d$ by 
$$
d:=\sqrt{b^2-4a_Nc_{N-1}X_1}
$$
and use the relation (\ref{a_(i,i-1)}) in the Appendix for $i=1$, we obtain
\begin{eqnarray*}
b&=&c_N-a_{N-1}X_1+2a_{N-1}X_1\\
&=&
\sum_{l=0}^{[N/2]}~ C_l
          +2a_{N-1}X_1,
\end{eqnarray*}
where
\begin{equation}
C_l:=C_l^{(1,N)}=(-1)^l~\sum_{j_1,j_2,\cdots,j_l}^{(1,N)} X_{j_1} X_{j_2} \cdots X_{j_l}.
\label{eq:N-cons}
\end{equation}
Moreover, using the result of Appendix (\ref{X...X}),
\begin{equation}
(c_Na_{N-1}-a_Nc_{N-1})X_1=X_1X_2\cdots X_N
\label{X_1...X_N}
\end{equation}
we find the expression for $d$
\begin{eqnarray}
 d = \sqrt{\left(\sum_{l=0}^{[N/2]} C_l\right)^2
         - 4X_1X_2 \cdots X_N},
 \label{eq:d}
\end{eqnarray}
which is symmetric under the permutation of $X_n$'s.
\\
We have arrived at the expression which looks complicated. The map could be single valued only if $d$ vanishes and could be regular only if the square root is resolved. How the equation could be completely integrable.
\\
A miracle happens if we substitute (\ref{X_n=}) into (\ref{eq:d}). The square root is resolved and we get (see Appendix)
\begin{eqnarray}
 d= \biggl| x_1x_2 \cdots x_N - (1-x_1)(1-x_2) \cdots (1-x_N) \biggl|.
\label{d=||}
\end{eqnarray}
At the same time we also observe that the two-valuedness of the solutions is inevitable. Therefore the orbits will split to two at every step of the map. The situation is just like it happend in the 3-point case, but the expression of solutions looks much more complicated. This is, however, not a problem, because we can find one solution out of two easily. Namely if we suppose a set of solutions as $x'_n=1-x_{n+1}$ for all $n$ and substitute into the equation (\ref{eq:Nd-LV simple}), we will immediately see that it is satisfied. Once we know one of solutions, say $x_-$, the other one is readily found to be $b/a_N\ -\ x_-$ from (\ref{x'_1}). In summary we have obtained 
\begin{equation}
 x'_1 = \left\{
 \begin{array}{l}
  \cfrac{b}{a_N}-1+x_2\\
  \\
  1-x_{2}
 \end{array}
 .
 \right.
 \label{eq:N-time-evo1}
\end{equation}
We obtain general solutions from (\ref{eq:N-time-evo1}) by symmetric permutation of arguments:
\begin{eqnarray}
 x'_n &=&
{a_{(n+1,n-1)}+a_{(n+2,n-2)}X_n\pm d\over 2a_{(n+2,n-1)}}\\
&=& \left\{
\begin{array}{c}
 \cfrac{a_{(n+1,n-1)}+a_{(n+2,n-2)}X_n}{a_{(n+2,n-1)}}-1+x_{n+1}\\
 \\
 1-x_{n+1}
\end{array}
\right.
\\
&=& \left\{
\begin{array}{c}
 x_n \cfrac{a_{(n+1,n-2)}}{a_{(n+2,n-1)}}
                   \hspace{10mm}  {\rm A-type} \\
 \\
 1-x_{n+1} \hspace{17mm} {\rm B-type}
\end{array}
\right.
.
\label{eq:N-time-evo}
\end{eqnarray}
To obtain the last expression, (\ref{mikan}) is used.
\\
The result we have obtained is quite similar with the case of three point d-LV equation. There are two types of distinct map at every step of time evolution. The B-type is a map such that $B^2$ is a cyclic permutation of points. When $N$ is even $B^N=1$, and when $N$ is odd $B^{2N}=1$ holds. We also find that $AB^2=B^2A$ is satisfied.
\\
\subsection{Conserved quantities}
We like to show, in the following, that the $C_l$'s which were defined in (\ref{eq:N-cons}) are constants of motion. To prove that we substitute (\ref{X_n=}) to (\ref{eq:N-cons}) and expand factors and then recombine terms of the same degree of powers. We obtain
\begin{eqnarray*}
&&\sum_{j_1,j_2,\cdots,j_l}^{(1,N)}x_{j_1}x_{j_2}\cdots x_{j_l}(1-x_{j_1+1})(1-x_{j_2+1})\cdots (1-x_{j_l+1})\\
&=&\sum_{r=0}^{l}(-1)^r\sum_{j_1,j_2,\cdots,j_l}^{(1,N)}x_{j_1}x_{j_2}\cdots x_{j_l}\sum_{a_1,a_2,\cdots,a_r}x_{j_{a_1}+1}x_{j_{a_2}+1}\cdots x_{j_{a_r}+1}\\
&=&\sum_{r=0}^{l}(-1)^r\sum_{j_1,j_2,\cdots,j_l}^{(1,N)}x_{j_1}x_{j_2}\cdots x_{j_l}\sum_{a_1,a_2,\cdots,a_r}x_{j_{a_1}-1}x_{j_{a_2}-1}\cdots x_{j_{a_r}-1}.
\end{eqnarray*}
To obtain the last expression we have used the fact that, owing to the cyclic symmetry of the summations, there are always the term $x_{j_{a_1}-1}x_{j_{a_2}-1}\cdots x_{j_{a_r}-1}$ in the summation if there is $x_{j_{a_1}+1}x_{j_{a_2}+1}\cdots x_{j_{a_r}+1}$. Then we can recombine terms to get
$$
\sum_{j_1,j_2,\cdots,j_l}^{(1,N)}x_{j_1}x_{j_2}\cdots x_{j_l}(1-x_{j_1-1})(1-x_{j_2-1})\cdots (1-x_{j_l-1}).
$$
Now by usig the equation of motion (\ref{eq:Nd-LV simple}), this is nothing but $\sum_{j_1,j_2,\cdots,j_l}X_{j_1}X_{j_2}\cdots X_{j_l}$ evaluated at time $t-1$. Therefore we conclude that $C_l$ with $l=1,2,\cdots,[N/2]$ are constants of motion.
\\
Now we consider relations of the two time-maps and the conserved quantities. To this end we first notice that $X_1X_2\cdots X_N$ is also conserved, as it is apparent from the equation of motion. Since $C_l$'s are constants, $d$ of (\ref{eq:d}) is also a constant. If we define
\begin{eqnarray}
 r:=x_1 x_2 \cdots x_N,\qquad 
 s:=(1-x_1)(1-x_2) \cdots (1-x_N),
\end{eqnarray}
we have 
$$
rs=X_1X_2\cdots X_N,\qquad |r-s|=d.
$$
Hence we find that $rs$ and $|r-s|$ are both constants of motion. 
\\
In the 3-point d-LV the B-type maps are characterized by the exchange of $r$ and $s$, while they remain unchanged by the A-type maps. This is true also for arbitrary $N$ point d-LV. For example we see, for the A-type map of (\ref{eq:N-time-evo}),
\begin{eqnarray*}
x'_1x'_2\cdots x'_N&=&x_1x_2\cdots x_N{a_{(2,-1)}\over a_{(3,0)}}
{a_{(3,0)}\over a_{(4,1)}}\cdots {a_{(N+1,N-2)}\over a_{(N+2,N-1)}}\\
&=&
x_1x_2\cdots x_N
\end{eqnarray*}
Since $rs$ is an apparent constant this result also means $s$ does not change. Therefore we conclude
\begin{itemize}
 \item $A$-type \hspace{5mm}  $r \rightarrow r ~~,~~ s
  \rightarrow s$~~~($r,s$ are unchanged)
 \item $B$-type \hspace{5mm}  $r \rightarrow s ~~,~~ s
  \rightarrow r$~~~($r,s$ are exchanged).
\end{itemize}

\subsection{Continuous time limit}
From the construction of d-LV equation there is no problem to obtain (\ref{eq:N-LV}) as the continuous time limit of (\ref{eq:Nd-LV}). As it was the case of 3-point d-LV, this does not mean that all solutions of d-LV remain being solutions of (\ref{eq:Nd-LV}) in the limit $\delta\rightarrow 0$.
\\
For the A-type map of (\ref{eq:N-time-evo}), collecting the terms of lowest order in $\delta$ we obtain
\begin{eqnarray}
 x^t_n + \delta \frac{dx_n^t}{dt} +o(\delta^2)
 = x^t_n \cfrac{1-\delta(x^t_{n+1}+\cdots+x^t_{n+N-2})+o(\delta^2)}
               {1-\delta(x^t_{n+2}+\cdots+x^t_{n+N-1})+o(\delta^2)}
\end{eqnarray}
or
\begin{eqnarray}
 \frac{dx_n^t}{dt}&=& x^t_n(x^t_{n+2}+\cdots+x^t_{n+N-2}+x^t_{n+N-1})
 \nonumber \\
 && -x^t_n(x^t_{n+1}+x^t_{n+2}+\cdots+x^t_{n+N-2})
 \nonumber \\
  &=& x_n^t(x_{n+1}^t-x_{n-1}^t).
\end{eqnarray}
Hence we find that the A-type map of (\ref{eq:N-time-evo}) reproduces (3), which means that every solution consisting of a sequence of A-type map is a solution of (\ref{eq:N-LV}) in the continuous time limit.

As for the B-type map of (\ref{eq:N-time-evo}) we have the expression 
$$
x^{t+\delta}_n={1\over\delta}-x^t_{n+1},
$$
after restoring $\delta$. All the solutions of this type tend to infinity in the limit $\delta\rightarrow 0$.

\section{Ultra discrete limit}

The ultra discrete Lotka-Volterra equation is another integrable deformation of the LV equation. Since it is derived starting from the d-LV equation which include both A-type map and B-type map, we are interested in how the solutions to d-LV equation behave in this limit. 

\subsection{3-point case}

The ultra discrete version of LV equation was shown to exist in \cite{TTMS}. 
3-point ud-LV equation under periodic condition is:
\begin{equation}
\left\{
\begin{array}{ccl}
 d'_1-d_1 &=& \max[0,d_2-1]-\max[0,d'_3-1],\\
 d'_2-d_2 &=& \max[0,d_3-1]-\max[0,d'_1-1],\\
 d'_3-d_3 &=& \max[0,d_1-1]-\max[0,d'_2-1],
\end{array}
\right.
\label{eq:3ud-LV}
\end{equation}
where $\max[a,b]$ means to choose $a$ if $a\ge b$ and $b$ if $b>a$.
This equation is derived from such transformation of 
variables:
\begin{eqnarray}
\delta=\exp[-1/\epsilon],\qquad x^t_n = -\exp (d^t_n/\epsilon)
\label{eq:tra-ud}
\end{eqnarray}
and ultra discrete limit\footnote{
When we derive (\ref{eq:3ud-LV}), we also use the property
\[ 
\lim_{\epsilon\rightarrow +0} \epsilon\log\left(1+e^{X/\epsilon}\right)
= \max[0,X] . \]
}
 $\epsilon\rightarrow +0$. \\
First let us attempt, like the d-LV case, to solve ud-LV equation (\ref{eq:3ud-LV}) for $(d'_1,d'_2,d'_3)$ in terms of $(d_1,d_2,d_3)$. Successive substitutions yield
\begin{eqnarray*}
 d'_1 &=& d_1 + \max[0,d_2-1]-\max[0,d'_3-1] \\
 &=& d_1 + \max[0,d_2-1] \\
 && -\max\Bigg[~0~,~d_3 + \max[0,d_1-1]-\max[0,d'_2-1] -1\Bigg] \\
 &=& d_1 + \max[0,d_2-1] \\
 && -\max\Bigg[~0~,~d_3 + \max[0,d_1-1] \label{eq:ud-maxmax}\\
 && -\max\biggl[~0~,~d_2 + \max[0,d_3-1]-\max[0,d'_1-1] -1\biggl] -1\Bigg],
\end{eqnarray*}
which is to be solved for $d'_1$. The appearance of $d'_1$ in $\max$ statement in the right hand side prevents us to proceed this calculas further. Thus we must seek other method to find solutions.
\\
For the meantime we suppose $(d_1,d_2,d_3)$ have a specific set of integers.
Then we can seek $(d'_1,d'_2,d'_3)$ following to the prossesses below:
\begin{enumerate}
 \item Suppose $d'_1$ is an integer $a$.
 \item From the value $a$ and the second relation of (\ref{eq:3ud-LV})  we determine $d'_2$.
 \item Similaly we can determine $d'_3$ from above $d'_2$ and the third relation of (\ref{eq:3ud-LV}).
 \item Check if these values $(d'_1,d'_2,d'_3)$ satisfy the first relation
       of (\ref{eq:3ud-LV}). If so,  
       $(d'_1,d'_2,d'_3)$ is a correct set of time evolution.
 \item Return to 1. and try another integer.
\end{enumerate}
One may think there are more than one correct set of time evolution.
But we find numerically only one correct set of $(d'_1,d'_2,d'_3)$ 
starting from a given set of $(d_1,d_2,d_3)$. We have checked this for every integer $a$ between $\pm 1000$. In this way we are convinced by ourselves that $(d'_1,d'_2,d'_3)$ is uniquely determined. 
\\
Figure \ref{fig:ex-ud} shows some examples of such ud-LV system.
\begin{figure}[ht]
\begin{center}
$
\begin{array}{lccc}
t= 0  &  (2 &  1 &  1 )\\ 
t= 1  &  (1 &  1 &  2 )\\ 
t= 2  &  (1 &  2 &  1 )\\ 
t= 3  &  (2 &  1 &  1 )\\ 
\end{array}
\hspace{2cm}
\begin{array}[bt]{lccc}
t= 0  &  (0 &  1 &  2) \\ 
t= 1  &  (0 &  2 &  1) \\ 
t= 2  &  (1 &  2 &  0) \\ 
t= 3  &  (2 &  1 &  0) \\ 
t= 4  &  (2 &  0 &  1) \\ 
t= 5  &  (1 &  0 &  2) \\ 
t= 6  &  (0 &  1 &  2) \\ 
\end{array}
\hspace{2cm}
\begin{array}[bt]{lccc}
t= 0  &  (8 &  0 &  4) \\
t= 1  &  (0 &  3 &  9) \\ 
t= 2  &  (2 & 10 &  0) \\ 
t= 3  & (11 &  0 &  1) \\ 
t= 4  &  (1 &  0 & 11) \\ 
t= 5  &  (0 & 10 &  2) \\ 
\cdots &   &   &   \\
t=33  &  (8 &  0 &  4) \\ 
\end{array}
$
\caption{Example of 3-Point ultra discrete Lotka-Volterra system}\label{fig:ex-ud}
\end{center}
\end{figure}

Now let us see ultra discrete limit of solutions.
First we will consider ultra discrete limit of A-type.
We again start from the A-type solution of (\ref{A type with delta}). By the transformation (\ref{eq:tra-ud}), it turns to
\begin{eqnarray}
 e^{\frac{1}{\epsilon}d'_1}=e^{\frac{1}{\epsilon}d_1}
 \cfrac{1+e^{\frac{1}{\epsilon}(d_2-1)}+e^{\frac{1}{\epsilon}(d_2+d_3-2)}}
       {1+e^{\frac{1}{\epsilon}(d_3-1)}+e^{\frac{1}{\epsilon}(d_3+d_1-2)}},
\end{eqnarray}
hence
\begin{eqnarray}
 d'_1-d_1=\epsilon\ln
 \left(1+e^{\frac{1}{\epsilon}(d_2-1)}+e^{\frac{1}{\epsilon}(d_2+d_3-2)}\right)-
 \epsilon\ln
 \left(1+e^{\frac{1}{\epsilon}(d_3-1)}+e^{\frac{1}{\epsilon}(d_3+d_1-2)}\right).
\end{eqnarray}
The ultra discrete limit $\epsilon\rightarrow +0$ yields
\begin{eqnarray}
 d'_1=d_1+\max[0,d_2-1,d_2+d_3-2]-\max[0,d_3-1,d_3+d_1-2].
 \label{eq:A-ud}
\end{eqnarray}
Since there is no $d'_1$ in $\max$ statement in the right hand side, we can get directly information for $d'_1$ from preceeding ones. The general formula for other $d_n$ can be found from (\ref{eq:A-ud}) as
\begin{eqnarray}
 d'_n=d_n+\max[0,d_{n+1}-1,d_{n+1}+d_{n-1}-2]-\max[0,d_{n-1}-1,d_{n-1}+d_n-2]
 \label{eq:A-ud for d_n}
\end{eqnarray}
From (\ref{eq:A-ud for d_n}) we find immediately the following results:
\begin{itemize}
 \item If all of $d_n$ are less than 1, all $d_n$ remain constant.
 \item If all of $d_n$ are greater than 1,
       $d'_n=d_{n-1}$.
\end{itemize}
We will have more complicated pattern of time evolution if there are $d_n$'s, some of which are less than 1 and some others are greater than 1.\\
As for the B-type map the solution in (\ref{A type with delta}) becomes, after the transformation (\ref{eq:tra-ud}),
\begin{eqnarray}
 1+e^{\frac{1}{\epsilon}(d'_1-1)}+e^{\frac{1}{\epsilon}(d_2-1)} = 0.
\end{eqnarray}
By taking logarithm of both sides and letting $\epsilon$ go to $+0$ we get
\begin{eqnarray}
 \lim_{\epsilon\rightarrow +0}\epsilon\ln\left(
 1+e^{\frac{1}{\epsilon}(d'_1-1)}+e^{\frac{1}{\epsilon}(d_2-1)}\right)&=&\max[0, d'_1-1, d_2-1]\\
&=& \epsilon\times(-\infty).
\end{eqnarray}
This is consistent for none of $d_n$'s. Therefore we conclude that the B-type map does not give any meaningful result in the ultra discrete limit.

\subsection{Generalization to N-point case}

$N$-point ultra discrete LV under periodic condition has been given by\cite{TTMS}
\begin{eqnarray}
 d^{t+1}_n - d^t_n
 = \max[0,d^{t}_{n+1}-1]-\max[0,d^{t+1}_{n-1}-1],~~~~
 n = 1,2,\cdots,N.
\label{eq:ud-lv2}
\end{eqnarray}
This can be derived from (\ref{eq:Nd-LV}) using the rule of transformations (\ref{eq:tra-ud}).
\\
By the same reason as in the 3-point ud-LV case we cannot derive simple ud-LV time-map directly from (\ref{eq:ud-lv2}), since the unknown variables are under the max statement. We can, however,  derive it from the A-type map (\ref{eq:N-time-evo}). According to the prescription of ultra discretization (\ref{eq:tra-ud}) and taking the ultra discrete limit
$\epsilon\rightarrow +0$, we have
$$
d'_n=d_n+\lim_{\epsilon\rightarrow 0}\left(\epsilon \ln a_{(n+1,n-2)}-\epsilon \ln a_{(n+2,n-1)}\right)
$$
The ultra discrete transformation of $X_j$ is given by
$$
X_j=\delta x_j(1-\delta x_{j+1})\quad\stackrel{ud}{\longrightarrow}\quad -\exp((d_j-1)/\epsilon)-\exp((d_j+d_{j+1}-2)/\epsilon)
$$
Since, in $a_{(n+1,n-2)}$ and $a_{(n+2,n-1)}$, $X_j$'s appear in the form $(-1)^lX_{j_1}X_{j_2}\cdots X_{j_l}$, 
$$
a_{(i,k)}\stackrel{ud}{\longrightarrow}\sum_{l=0}^{[(|k-i|+2)/2]}\sum^{(i,k)}_{{j_1},{j_2},\cdots,{j_l}}
\exp\left((d_{j_1}+d_{j_2}+\cdots +d_{j_l}-l)/\epsilon\right)
$$
$$
\times \left\{1+\exp((d_{j_1+1}-1)/\epsilon)+\exp((d_{j_2+1}-1)/\epsilon)+\cdots +\exp((d_{j_l+1}-1)/\epsilon)\right.
$$
$$
\left.+\exp((d_{j_1+1}+d_{j_2+1}-2)/\epsilon)+\cdots+
\exp((d_{j_1+1}+d_{j_2+1}+\cdots+d_{j_l+1}-l)/\epsilon)\right\}.
$$
Let us denote by $(a,b)\oplus(c,d)$ the set $(a+c,a+d,b+c,b+d)$. Then we have
$$
\lim_{\epsilon\rightarrow 0}\epsilon (\ln\ a_{(i,k)})
=
\max\left[\bigcup_{l=0}^{[(|k-i|+2)/2]}\bigcup^{(i,k)}_{{j_1},{j_2},\cdots,{j_l}}\oplus\left(m_{j_1}\oplus m_{j_2}\oplus \cdots \oplus m_{j_l}\right)\right]
$$
where
$$
m_{j}:=(d_{j}-1,\ d_{j}+d_{j+1}-2).
$$
Finally we obtain the solution of ud-LV as
\begin{eqnarray*}
d'_n=d_n&+&\max\left[\bigcup_{l=0}^{[(N-1)/2]}\bigcup^{(n+1,n-2)}_{{j_1},{j_2},\cdots,{j_l}}\oplus\left(m_{j_1}\oplus m_{j_2}\oplus \cdots \oplus m_{j_l}\right)\right]\\
&-&\max\left[\bigcup_{l=0}^{[(N-1)/2]}\bigcup^{(n+2,n-1)}_{{j_1},{j_2},\cdots,{j_l}}\oplus\left(m_{j_1}\oplus m_{j_2}\oplus \cdots \oplus m_{j_l}\right)\right].
\end{eqnarray*}
This map will give the same map derived from (\ref{eq:ud-lv2}). It, however, comes only from the A-type map. As for the B-type map we will not get a meaningful answer by the same reason we discussed in the case of 3-point ud-LV equation. We could guess this result because the ultra discrete limit also implies the continuous time limit as we could see from (\ref{eq:tra-ud}).
\\
\section{Concluding remarks}

In this paper we have discussed properties of discrete Lotka-Volterra equation. A special attention has been focussed on the fact that the map is not single valued at every step of the time evolution. This singular behaviour disappears both in the continuous time and ultra discrete limits, and uniqueness of the map is fulfilled.
\\
The violation of the uniqueness of the d-LV system does not contradict with complete integrability, in the sense that it has sufficient number of constants of motion. A singular map, which we called B-type, does not create new orbits during the evolution but simply causes jumps of the system between two possible orbits. 
\\
We have explained in detail how these two maps arise in the d-LV system. Usually such a solution as B-type was not taken into account to study a deformation of the continuous time Lotka-Volterra equation\cite{NH}. In principle, however, there is no a priopi reason to ignore the B-type solutions. 

General solutions are random sequence of A-type and B-type maps. In other words two types of solutions can be chosen arbitrarily at every step of the map, hence we can not write their orbits in a compact form. If only A-type of map is selected every time many solutions are known explicitly by means of the $\tau$ functions of the theory of Toda lattice. We will present here a particular one of such solution which corresponds to the cnoidal wave solution of the continuous time Toda lattice, i.e., 
\begin{eqnarray}
 V_n(t) = (2K\nu)^2 \biggl[\mbox{dn}^2 
 \Bigl\{2(\nu t\pm\frac{n}{\lambda})K \Bigl\}-\frac{E}{K}\biggl]
\label{cnoidal}
\end{eqnarray}
where dn is a Jacobi elliptic function, $K$ and $E$ are called 1st and 2nd complete elliptic integrals, respectively, and $\nu$ and $\lambda$ are related by the spectral relation
\begin{eqnarray}
 (2K\nu)^2 = \biggl\{
 \frac{1}{\mbox{sn}^2(2K/{\lambda})}-1+\frac{E}{K}\biggl\}^{-1}.
\end{eqnarray}

In terms of $\tau$ functions the function $V_n$ of the Toda lattice is expressed as
\begin{equation}
V_n(t)= \frac{{\tau}_{n-1}^{t+1}{\tau}_{n+1}^{t}}
                {{\tau}_n^{t+1}{\tau}_n^t} 
\label{V-tau}
\end{equation}
hence, by means of the Miura transformation (\ref{eq:Miura-T2}), the function of d-LV is given by
\begin{eqnarray}
x_n^t = \frac{\tau^{t+1}_{(n-2)/2}\tau^t_{(n+1)/2}}
                {\tau^{t+1}_{(n-1)/2}\tau^t_{n/2}}.
\label{x-tau}
\end{eqnarray}
Corresponding to the cnoidal wave (\ref{cnoidal}) $\tau_n^t$ is given by
\begin{eqnarray}
 {\tau}_n^t
 = \vartheta_0\Bigl(\nu t \pm\frac{n}{{\lambda}}+\phi\Bigl)
\end{eqnarray}
where $\vartheta_0$ is the Jacobi theta function and $\phi$ is an arbitrary constant phase. Substitution of this into (\ref{x-tau}) yields
\begin{eqnarray}
 x_n^t
  &=&\cfrac{\vartheta_0(\nu(t+1) \pm\frac{n-1}{2\lambda}+{\phi})
           \vartheta_0(\nu t \pm\frac{n+2}{2\lambda}+{\phi})}
          {\vartheta_0(\nu(t+1) \pm \frac{n-1}{2\lambda}+{\phi})
           \vartheta_0(\nu t \pm \frac{n}{2\lambda}+{\phi})}\\
 &=& A~ \cfrac{1-k^2 \mbox{sn}^2\xi~ \mbox{sn}^2\eta}
             {1-k^2 \mbox{sn}^2\xi~ \mbox{sn}^2\tilde{\eta}}
\label{cnoidal x_n}
\end{eqnarray}
where
\begin{eqnarray}
 \xi := 2K\left(\nu t \pm \frac{n}{2\lambda}+\tilde\phi\right) &,&
 \eta = K\left({\nu}\mp\frac{3}{2\lambda}\right) ~,~
 \tilde{\eta} = K\left(\nu\mp{1\over 2\lambda}\right) , \\
 A = \frac{\vartheta_0^2(\eta/2K)}{\vartheta_0^2(\tilde{\eta}/2K)} ~
  &,&
 \tilde\phi={\phi}+\frac{\nu}{2}\mp\frac{1}{4\lambda}
\end{eqnarray}
and $k$ is the modulas which characterizes the elliptic functions. 
We have used
$$
1-k^2 \mbox{sn}^2\xi~ \mbox{sn}^2\eta={\vartheta^2_0(0)\vartheta_0(x+y)\vartheta(x-y)\over \vartheta^2_0(x)\vartheta^2_0(y)},\quad
\xi=2Kx,\quad \eta=2Ky .
$$

The Jacobi sn function has a symmetry
$$
{\rm sn} (\xi+2K)=-{\rm sn}\xi.
$$
Comparing (\ref{cnoidal x_n}) with (\ref{eq:period1}) the periodicity condision on $x_n^t$ requires for $\lambda$ to satisfy
$$
\lambda={N\over l},\qquad l=1,2,3,\cdots.
$$
In the case of three point d-LV, $\lambda=3,3/2,1,3/4,\cdots$.

If only B-type of map is allowed to exist, we have general solution in the following form:
\begin{equation}
 x^t_n = \left\{
  \begin{array}{c}
   f(n+t) \hspace{20mm} \mbox{if}~t~odd \\
   \\
   \cfrac{1}{\delta}-f(n+t) \hspace{12mm} \mbox{if}~t~even
  \end{array}
  \right.
\end{equation}
where $f(m)$ is an arbitrary function of period $N$, i.e., $f(m+N)=f(m)$.
\\

It will be interesting to understand what is the corresponding behaviour of the B-type map in the discrete time Toda lattice. The answer is rather simple. It corresponds to exchange the role of $V$ and $I$, as we can see directly in the Miura transformation (\ref{eq:Miura-T2}). This property of the Toda lattice is transformed into two distinct orbits, which we called $O_\pm$, of the d-LV system. We note that this symmetry is different from what is called duality under the $V-I$ exchange in the case of continuous time \cite{Toda}, since our symmetry arises only in the discrete time evolution.

Before closing this paper we like to point out that the singular behaviour of the map, which we discussed, is not special in d-LV equation. As we will explain in the other paper, this nonuniqueness of the map turns out to be a source of Julia set which should be generated when the equation of motion is deformed from the integrable case. In order to make this point more clear we recall how the integrable logistic map is changed to the non-integrable one. 
\\
In a series of our previous papers\cite{SSKY} we have studied a generalization of logistic map which interpolates between the M\"obius map and the logistic map. The former is integrable whereas the latter is well known non-integrable map which generates chaos. To be specific we consider
\begin{equation}
z_{t+1}=\mu z_t{1-\gamma z_t\over 1+\mu(1-\gamma)z_t}.
\label{glm}
\end{equation}
$\gamma=0$ and $\gamma=1$ are the M\"obius and the logistic map respectively. The M\"obius map can be integrated explicitly as
\begin{equation}
z_t=\cfrac{\mu^t}{a+\mu\cfrac{1-\mu^t}{1-\mu}}.
\end{equation}
If $\gamma\ne 0$, the map (\ref{glm}) is not integrable. Depending on its initial values as well as on the values of parameters $\mu$ and $\gamma$, orbits converge to some attractors, otherwise behave chaotic. In order to see if the map is integrable or not, it is more appropriate to study Julia set of the map. The Julia set is defined on the complex plane of $z_t$. It is a closure of repulsive fixed points, which can be obtained by the inverse map starting from an arbitrary repulsive fixed point. Since it is an invariant set of the map it does not depend on which repulsive point is chosen at first. There exists a Julia set on the complex plane if the map is not integrable. Therefore we are interested in how the Julia set will disappear in the integrable limit $\gamma\rightarrow 0$.
\\
The inverse map of (\ref{glm}) is readily solved 
\begin{equation}
z_t=\cfrac{1-(1-\gamma)z_{t+1}\pm\sqrt{(1-(1-\gamma)z_{t+1})^2-4\gamma z_{t+1}/\mu}}{2\gamma}.
\label{pm}
\end{equation}
This map is not unique and the Julia set has a fractal structure. In the integrable limit $\gamma\rightarrow 0$ we obtain
\begin{equation}
z_t=\left\{
\begin{array}{c}
 2\mu(1-z_{t+1})\\
 0
\end{array}
\right.
\end{equation}
hence the square root singularity is resolved but the map is not unique. The `B-type' map does not create new orbits. It maps every point to the origin. If we produce the Julia set by using computer we will not observe this point. But it is not right to forget this point. When the equation is deformed from the integrable one two maps of (\ref{pm}) are equal footing and become the source of chaos. The situation is exactly the same with what we learned in this and the following papers.
\vglue 1cm

\vglue 1cm
\noindent
{\bf Acknowledgements}

This work is supported in part by the Grant-in-Aid for general Scientific Research from the Ministry of Education, Sciences, Sports and Culture, Japan (No 10640278).

\appendix
\section*{Appendix}
Let us start from the definition (\ref{a_(i,k)}):
\[
a_{(i,k)}:=\sum_{l=0}^{l_{max}}C_l^{(i,k)},\qquad 
l_{max}=\left\{ \begin{array}{ll}
                     [(k-i+2)/2]  & k\ge i \\
                    {[(N-i+k+2)/2]} & k<i
                \end{array}\right.
\]
This satisfies the recurrence formulae
\begin{eqnarray}
a_{(i-1,k+1)}&=&a_{(i-1,k)}-a_{(i-1,k-1)}X_{k+1},
\label{a_(i-1,k+1)=a_(i-1,k)}\\
&=&a_{(i,k+1)}-a_{(i+1,k+1)}X_{i-1},
\label{a_(i-1,k+1)=a_(i,k+1)}
\end{eqnarray}
when $k-i\ne -1,-2\ mod \  N$, and
\begin{eqnarray}
a_{(i,i-1)}&=&
a_{(i+1,i-1)}-a_{(i+2,i-2)}X_{i},\nonumber\\
&=&
a_{(i,i-2)}-a_{(i+1,i-3)}X_{i-1},
\label{a_(i,i-1)},
\end{eqnarray}
when $k-i= -1\ mod \  N$.

Using (\ref{a_(i-1,k+1)=a_(i,k+1)})
\begin{eqnarray}
A_{i,k}&:=&a_{(i,k)}-(1-x_{i})a_{(i+1,k)}\nonumber\\
&=&
a_{(i+1,k)}-a_{(i+2,k)}X_i-(1-x_{i})a_{(i+1,k)}\nonumber\\
&=&x_iA_{i+1,k}=\cdots\nonumber\\
&=&x_i x_{i+1}\cdots x_{k+1}.
\label{A_ik}
\end{eqnarray}
Similarly using (\ref{a_(i-1,k+1)=a_(i-1,k)}),
\begin{eqnarray}
B_{i,k}&:=&a_{(i,k)}-x_{k+1}a_{(i,k-1)}\nonumber\\
&=&
a_{(i,k-1)}-a_{(i,k-2)}X_k-x_{k+1}a_{(i,k-1)}\nonumber\\
&=&B_{i,k-1}(1-x_{k+1})=\cdots\nonumber\\
&=&(1-x_{i})\cdots (1-x_{k})(1-x_{k+1}).
\end{eqnarray}

Let us prove (\ref{d=||}). Using (\ref{a_(i,i-1)})
\begin{eqnarray}
\sum_{l=0}^{[N/2]}~ C_l
&=&
a_{(i,i-1)}=a_{(i,i-2)}-a_{(i+1,i-3)}X_{i-1}\nonumber\\
&=&
B_{i,i-3}(1-x_{i-1})+A_{i,i-3}x_{i-1}\nonumber\\
&=&
\prod_{j=1}^N(1-x_j)+\prod_{j=1}^Nx_j.
\label{prod(1-x_j)+prod x_j}
\end{eqnarray}

(\ref{X_1...X_N}) can be shown by using (\ref{a_(i-1,k+1)=a_(i-1,k)}) as
\begin{eqnarray}
a_{(i-1,k+1)}a_{(i,k)}-a_{(i-1,k)}a_{(i,k+1)}
&=&\left(a_{(i-1,k)}a_{(i,k-1)}-a_{(i-1,k-1)}a_{(i,k)}\right)X_{k+1}\nonumber\\
&=&\cdots\nonumber\\
&=&-X_{i-1}X_i\cdots X_{k+1}
\label{X...X}
\end{eqnarray}

Finally we prove (\ref{eq:N-time-evo}):
\begin{eqnarray}
&&(x_{n+1}-1)a_{(n+2,n-1)}+a_{(n+1,n-1)}+a_{(n+2,n-2)}X_n
\nonumber\\
&=&
A_{n+1,n-1}-A_{n+1,n-2}x_n+a_{(n+1,n-2)}x_n\nonumber\\
&=&
a_{(n+1,n-2)}x_n
\label{mikan}
\end{eqnarray}
where (\ref{A_ik}) is used.


\end{document}

%% file: hozon.tex

\setlength{\unitlength}{0.240900pt}
\ifx\plotpoint\undefined\newsavebox{\plotpoint}\fi
\sbox{\plotpoint}{\rule[-0.200pt]{0.400pt}{0.400pt}}%
\begin{picture}(1200,900)(0,0)
\font\gnuplot=cmr10 at 10pt
\gnuplot
\sbox{\plotpoint}{\rule[-0.200pt]{0.400pt}{0.400pt}}%
\put(220.0,113.0){\rule[-0.200pt]{220.664pt}{0.400pt}}
\put(220.0,113.0){\rule[-0.200pt]{0.400pt}{184.048pt}}
\put(220.0,113.0){\rule[-0.200pt]{4.818pt}{0.400pt}}
\put(198,113){\makebox(0,0)[r]{0}}
\put(1116.0,113.0){\rule[-0.200pt]{4.818pt}{0.400pt}}
\put(220.0,231.0){\rule[-0.200pt]{4.818pt}{0.400pt}}
\put(198,231){\makebox(0,0)[r]{0.2}}
\put(1116.0,231.0){\rule[-0.200pt]{4.818pt}{0.400pt}}
\put(220.0,348.0){\rule[-0.200pt]{4.818pt}{0.400pt}}
\put(198,348){\makebox(0,0)[r]{0.4}}
\put(1116.0,348.0){\rule[-0.200pt]{4.818pt}{0.400pt}}
\put(220.0,466.0){\rule[-0.200pt]{4.818pt}{0.400pt}}
\put(198,466){\makebox(0,0)[r]{0.6}}
\put(1116.0,466.0){\rule[-0.200pt]{4.818pt}{0.400pt}}
\put(220.0,583.0){\rule[-0.200pt]{4.818pt}{0.400pt}}
\put(198,583){\makebox(0,0)[r]{0.8}}
\put(1116.0,583.0){\rule[-0.200pt]{4.818pt}{0.400pt}}
\put(220.0,701.0){\rule[-0.200pt]{4.818pt}{0.400pt}}
\put(198,701){\makebox(0,0)[r]{1}}
\put(1116.0,701.0){\rule[-0.200pt]{4.818pt}{0.400pt}}
\put(220.0,818.0){\rule[-0.200pt]{4.818pt}{0.400pt}}
\put(198,818){\makebox(0,0)[r]{1.2}}
\put(1116.0,818.0){\rule[-0.200pt]{4.818pt}{0.400pt}}
\put(220.0,113.0){\rule[-0.200pt]{0.400pt}{4.818pt}}
\put(220,68){\makebox(0,0){0}}
\put(220.0,857.0){\rule[-0.200pt]{0.400pt}{4.818pt}}
\put(403.0,113.0){\rule[-0.200pt]{0.400pt}{4.818pt}}
\put(403,68){\makebox(0,0){0.2}}
\put(403.0,857.0){\rule[-0.200pt]{0.400pt}{4.818pt}}
\put(586.0,113.0){\rule[-0.200pt]{0.400pt}{4.818pt}}
\put(586,68){\makebox(0,0){0.4}}
\put(586.0,857.0){\rule[-0.200pt]{0.400pt}{4.818pt}}
\put(770.0,113.0){\rule[-0.200pt]{0.400pt}{4.818pt}}
\put(770,68){\makebox(0,0){0.6}}
\put(770.0,857.0){\rule[-0.200pt]{0.400pt}{4.818pt}}
\put(953.0,113.0){\rule[-0.200pt]{0.400pt}{4.818pt}}
\put(953,68){\makebox(0,0){0.8}}
\put(953.0,857.0){\rule[-0.200pt]{0.400pt}{4.818pt}}
\put(1136.0,113.0){\rule[-0.200pt]{0.400pt}{4.818pt}}
\put(1136,68){\makebox(0,0){1}}
\put(1136.0,857.0){\rule[-0.200pt]{0.400pt}{4.818pt}}
\put(220.0,113.0){\rule[-0.200pt]{220.664pt}{0.400pt}}
\put(1136.0,113.0){\rule[-0.200pt]{0.400pt}{184.048pt}}
\put(220.0,877.0){\rule[-0.200pt]{220.664pt}{0.400pt}}
\put(45,495){\makebox(0,0){Y}}
\put(678,23){\makebox(0,0){X}}
\put(220.0,113.0){\rule[-0.200pt]{0.400pt}{184.048pt}}
\put(1006,812){\makebox(0,0)[r]{'$O_+$'}}
\put(1050,812){\raisebox{-.8pt}{\makebox(0,0){$\diamond$}}}
\put(275,466){\raisebox{-.8pt}{\makebox(0,0){$\diamond$}}}
\put(337,473){\raisebox{-.8pt}{\makebox(0,0){$\diamond$}}}
\put(464,440){\raisebox{-.8pt}{\makebox(0,0){$\diamond$}}}
\put(625,361){\raisebox{-.8pt}{\makebox(0,0){$\diamond$}}}
\put(739,258){\raisebox{-.8pt}{\makebox(0,0){$\diamond$}}}
\put(782,182){\raisebox{-.8pt}{\makebox(0,0){$\diamond$}}}
\put(764,146){\raisebox{-.8pt}{\makebox(0,0){$\diamond$}}}
\put(679,133){\raisebox{-.8pt}{\makebox(0,0){$\diamond$}}}
\put(528,134){\raisebox{-.8pt}{\makebox(0,0){$\diamond$}}}
\put(380,147){\raisebox{-.8pt}{\makebox(0,0){$\diamond$}}}
\put(293,186){\raisebox{-.8pt}{\makebox(0,0){$\diamond$}}}
\put(258,266){\raisebox{-.8pt}{\makebox(0,0){$\diamond$}}}
\put(250,370){\raisebox{-.8pt}{\makebox(0,0){$\diamond$}}}
\put(259,445){\raisebox{-.8pt}{\makebox(0,0){$\diamond$}}}
\put(296,473){\raisebox{-.8pt}{\makebox(0,0){$\diamond$}}}
\put(386,463){\raisebox{-.8pt}{\makebox(0,0){$\diamond$}}}
\put(537,410){\raisebox{-.8pt}{\makebox(0,0){$\diamond$}}}
\put(686,314){\raisebox{-.8pt}{\makebox(0,0){$\diamond$}}}
\put(767,218){\raisebox{-.8pt}{\makebox(0,0){$\diamond$}}}
\put(781,161){\raisebox{-.8pt}{\makebox(0,0){$\diamond$}}}
\put(735,138){\raisebox{-.8pt}{\makebox(0,0){$\diamond$}}}
\put(616,132){\raisebox{-.8pt}{\makebox(0,0){$\diamond$}}}
\put(455,138){\raisebox{-.8pt}{\makebox(0,0){$\diamond$}}}
\put(332,161){\raisebox{-.8pt}{\makebox(0,0){$\diamond$}}}
\put(273,217){\raisebox{-.8pt}{\makebox(0,0){$\diamond$}}}
\put(252,313){\raisebox{-.8pt}{\makebox(0,0){$\diamond$}}}
\put(252,409){\raisebox{-.8pt}{\makebox(0,0){$\diamond$}}}
\put(271,463){\raisebox{-.8pt}{\makebox(0,0){$\diamond$}}}
\put(329,473){\raisebox{-.8pt}{\makebox(0,0){$\diamond$}}}
\put(449,445){\raisebox{-.8pt}{\makebox(0,0){$\diamond$}}}
\put(610,370){\raisebox{-.8pt}{\makebox(0,0){$\diamond$}}}
\put(1006,767){\makebox(0,0)[r]{'$O_-$'}}
\put(1050,767){\makebox(0,0){$\bullet$}}
\put(1081,348){\makebox(0,0){$\bullet$}}
\put(663,506){\makebox(0,0){$\bullet$}}
\put(681,680){\makebox(0,0){$\bullet$}}
\put(1086,589){\makebox(0,0){$\bullet$}}
\put(1056,340){\makebox(0,0){$\bullet$}}
\put(614,559){\makebox(0,0){$\bullet$}}
\put(755,681){\makebox(0,0){$\bullet$}}
\put(1099,541){\makebox(0,0){$\bullet$}}
\put(1015,342){\makebox(0,0){$\bullet$}}
\put(585,603){\makebox(0,0){$\bullet$}}
\put(839,679){\makebox(0,0){$\bullet$}}
\put(1105,487){\makebox(0,0){$\bullet$}}
\put(957,354){\makebox(0,0){$\bullet$}}
\put(573,636){\makebox(0,0){$\bullet$}}
\put(920,674){\makebox(0,0){$\bullet$}}
\put(1106,435){\makebox(0,0){$\bullet$}}
\put(882,377){\makebox(0,0){$\bullet$}}
\put(577,657){\makebox(0,0){$\bullet$}}
\put(988,664){\makebox(0,0){$\bullet$}}
\put(1102,392){\makebox(0,0){$\bullet$}}
\put(797,414){\makebox(0,0){$\bullet$}}
\put(597,670){\makebox(0,0){$\bullet$}}
\put(1037,646){\makebox(0,0){$\bullet$}}
\put(1094,363){\makebox(0,0){$\bullet$}}
\put(717,463){\makebox(0,0){$\bullet$}}
\put(635,677){\makebox(0,0){$\bullet$}}
\put(1070,619){\makebox(0,0){$\bullet$}}
\put(1077,346){\makebox(0,0){$\bullet$}}
\put(652,517){\makebox(0,0){$\bullet$}}
\put(694,681){\makebox(0,0){$\bullet$}}
\put(1089,580){\makebox(0,0){$\bullet$}}
\end{picture}

%% file: del.tex

\setlength{\unitlength}{0.240900pt}
\ifx\plotpoint\undefined\newsavebox{\plotpoint}\fi
\sbox{\plotpoint}{\rule[-0.200pt]{0.400pt}{0.400pt}}%
\begin{picture}(1200,900)(0,0)
\font\gnuplot=cmr10 at 10pt
\gnuplot
\sbox{\plotpoint}{\rule[-0.200pt]{0.400pt}{0.400pt}}%
\put(220.0,113.0){\rule[-0.200pt]{220.664pt}{0.400pt}}
\put(220.0,113.0){\rule[-0.200pt]{0.400pt}{184.048pt}}
\put(220.0,113.0){\rule[-0.200pt]{4.818pt}{0.400pt}}
\put(198,113){\makebox(0,0)[r]{0}}
\put(1116.0,113.0){\rule[-0.200pt]{4.818pt}{0.400pt}}
\put(220.0,287.0){\rule[-0.200pt]{4.818pt}{0.400pt}}
\put(198,287){\makebox(0,0)[r]{0.5}}
\put(1116.0,287.0){\rule[-0.200pt]{4.818pt}{0.400pt}}
\put(220.0,460.0){\rule[-0.200pt]{4.818pt}{0.400pt}}
\put(198,460){\makebox(0,0)[r]{1}}
\put(1116.0,460.0){\rule[-0.200pt]{4.818pt}{0.400pt}}
\put(220.0,634.0){\rule[-0.200pt]{4.818pt}{0.400pt}}
\put(198,634){\makebox(0,0)[r]{1.5}}
\put(1116.0,634.0){\rule[-0.200pt]{4.818pt}{0.400pt}}
\put(220.0,808.0){\rule[-0.200pt]{4.818pt}{0.400pt}}
\put(198,808){\makebox(0,0)[r]{2}}
\put(1116.0,808.0){\rule[-0.200pt]{4.818pt}{0.400pt}}
\put(220.0,113.0){\rule[-0.200pt]{0.400pt}{4.818pt}}
\put(220,68){\makebox(0,0){0}}
\put(220.0,857.0){\rule[-0.200pt]{0.400pt}{4.818pt}}
\put(312.0,113.0){\rule[-0.200pt]{0.400pt}{4.818pt}}
\put(312,68){\makebox(0,0){0.2}}
\put(312.0,857.0){\rule[-0.200pt]{0.400pt}{4.818pt}}
\put(403.0,113.0){\rule[-0.200pt]{0.400pt}{4.818pt}}
\put(403,68){\makebox(0,0){0.4}}
\put(403.0,857.0){\rule[-0.200pt]{0.400pt}{4.818pt}}
\put(495.0,113.0){\rule[-0.200pt]{0.400pt}{4.818pt}}
\put(495,68){\makebox(0,0){0.6}}
\put(495.0,857.0){\rule[-0.200pt]{0.400pt}{4.818pt}}
\put(586.0,113.0){\rule[-0.200pt]{0.400pt}{4.818pt}}
\put(586,68){\makebox(0,0){0.8}}
\put(586.0,857.0){\rule[-0.200pt]{0.400pt}{4.818pt}}
\put(678.0,113.0){\rule[-0.200pt]{0.400pt}{4.818pt}}
\put(678,68){\makebox(0,0){1}}
\put(678.0,857.0){\rule[-0.200pt]{0.400pt}{4.818pt}}
\put(770.0,113.0){\rule[-0.200pt]{0.400pt}{4.818pt}}
\put(770,68){\makebox(0,0){1.2}}
\put(770.0,857.0){\rule[-0.200pt]{0.400pt}{4.818pt}}
\put(861.0,113.0){\rule[-0.200pt]{0.400pt}{4.818pt}}
\put(861,68){\makebox(0,0){1.4}}
\put(861.0,857.0){\rule[-0.200pt]{0.400pt}{4.818pt}}
\put(953.0,113.0){\rule[-0.200pt]{0.400pt}{4.818pt}}
\put(953,68){\makebox(0,0){1.6}}
\put(953.0,857.0){\rule[-0.200pt]{0.400pt}{4.818pt}}
\put(1044.0,113.0){\rule[-0.200pt]{0.400pt}{4.818pt}}
\put(1044,68){\makebox(0,0){1.8}}
\put(1044.0,857.0){\rule[-0.200pt]{0.400pt}{4.818pt}}
\put(1136.0,113.0){\rule[-0.200pt]{0.400pt}{4.818pt}}
\put(1136,68){\makebox(0,0){2}}
\put(1136.0,857.0){\rule[-0.200pt]{0.400pt}{4.818pt}}
\put(220.0,113.0){\rule[-0.200pt]{220.664pt}{0.400pt}}
\put(1136.0,113.0){\rule[-0.200pt]{0.400pt}{184.048pt}}
\put(220.0,877.0){\rule[-0.200pt]{220.664pt}{0.400pt}}
\put(45,495){\makebox(0,0){Y}}
\put(678,23){\makebox(0,0){X}}
\put(220.0,113.0){\rule[-0.200pt]{0.400pt}{184.048pt}}
\put(1006,812){\makebox(0,0)[r]{'1.0'}}
\put(1050,812){\raisebox{-.8pt}{\makebox(0,0){$\diamond$}}}
\put(247,321){\raisebox{-.8pt}{\makebox(0,0){$\diamond$}}}
\put(279,325){\raisebox{-.8pt}{\makebox(0,0){$\diamond$}}}
\put(342,306){\raisebox{-.8pt}{\makebox(0,0){$\diamond$}}}
\put(423,446){\raisebox{-.8pt}{\makebox(0,0){$\diamond$}}}
\put(239,206){\raisebox{-.8pt}{\makebox(0,0){$\diamond$}}}
\put(556,267){\raisebox{-.8pt}{\makebox(0,0){$\diamond$}}}
\put(398,433){\raisebox{-.8pt}{\makebox(0,0){$\diamond$}}}
\put(256,158){\raisebox{-.8pt}{\makebox(0,0){$\diamond$}}}
\put(239,206){\raisebox{-.8pt}{\makebox(0,0){$\diamond$}}}
\put(556,267){\raisebox{-.8pt}{\makebox(0,0){$\diamond$}}}
\put(475,127){\raisebox{-.8pt}{\makebox(0,0){$\diamond$}}}
\put(413,124){\raisebox{-.8pt}{\makebox(0,0){$\diamond$}}}
\put(663,307){\raisebox{-.8pt}{\makebox(0,0){$\diamond$}}}
\put(422,260){\raisebox{-.8pt}{\makebox(0,0){$\diamond$}}}
\put(485,449){\raisebox{-.8pt}{\makebox(0,0){$\diamond$}}}
\put(235,266){\raisebox{-.8pt}{\makebox(0,0){$\diamond$}}}
\put(240,310){\raisebox{-.8pt}{\makebox(0,0){$\diamond$}}}
\put(259,326){\raisebox{-.8pt}{\makebox(0,0){$\diamond$}}}
\put(397,420){\raisebox{-.8pt}{\makebox(0,0){$\diamond$}}}
\put(273,143){\raisebox{-.8pt}{\makebox(0,0){$\diamond$}}}
\put(639,247){\raisebox{-.8pt}{\makebox(0,0){$\diamond$}}}
\put(418,375){\raisebox{-.8pt}{\makebox(0,0){$\diamond$}}}
\put(485,449){\raisebox{-.8pt}{\makebox(0,0){$\diamond$}}}
\put(235,266){\raisebox{-.8pt}{\makebox(0,0){$\diamond$}}}
\put(240,310){\raisebox{-.8pt}{\makebox(0,0){$\diamond$}}}
\put(418,375){\raisebox{-.8pt}{\makebox(0,0){$\diamond$}}}
\put(485,449){\raisebox{-.8pt}{\makebox(0,0){$\diamond$}}}
\put(659,368){\raisebox{-.8pt}{\makebox(0,0){$\diamond$}}}
\put(619,248){\raisebox{-.8pt}{\makebox(0,0){$\diamond$}}}
\put(500,140){\raisebox{-.8pt}{\makebox(0,0){$\diamond$}}}
\put(642,416){\raisebox{-.8pt}{\makebox(0,0){$\diamond$}}}
\put(651,252){\raisebox{-.8pt}{\makebox(0,0){$\diamond$}}}
\put(495,172){\raisebox{-.8pt}{\makebox(0,0){$\diamond$}}}
\put(600,439){\raisebox{-.8pt}{\makebox(0,0){$\diamond$}}}
\put(247,321){\raisebox{-.8pt}{\makebox(0,0){$\diamond$}}}
\put(279,325){\raisebox{-.8pt}{\makebox(0,0){$\diamond$}}}
\put(342,306){\raisebox{-.8pt}{\makebox(0,0){$\diamond$}}}
\put(422,260){\raisebox{-.8pt}{\makebox(0,0){$\diamond$}}}
\put(485,449){\raisebox{-.8pt}{\makebox(0,0){$\diamond$}}}
\put(235,266){\raisebox{-.8pt}{\makebox(0,0){$\diamond$}}}
\put(476,314){\raisebox{-.8pt}{\makebox(0,0){$\diamond$}}}
\put(423,446){\raisebox{-.8pt}{\makebox(0,0){$\diamond$}}}
\put(239,206){\raisebox{-.8pt}{\makebox(0,0){$\diamond$}}}
\put(235,266){\raisebox{-.8pt}{\makebox(0,0){$\diamond$}}}
\put(240,310){\raisebox{-.8pt}{\makebox(0,0){$\diamond$}}}
\put(259,326){\raisebox{-.8pt}{\makebox(0,0){$\diamond$}}}
\put(397,420){\raisebox{-.8pt}{\makebox(0,0){$\diamond$}}}
\put(565,445){\raisebox{-.8pt}{\makebox(0,0){$\diamond$}}}
\put(240,310){\raisebox{-.8pt}{\makebox(0,0){$\diamond$}}}
\put(259,326){\raisebox{-.8pt}{\makebox(0,0){$\diamond$}}}
\put(397,420){\raisebox{-.8pt}{\makebox(0,0){$\diamond$}}}
\put(565,445){\raisebox{-.8pt}{\makebox(0,0){$\diamond$}}}
\put(240,310){\raisebox{-.8pt}{\makebox(0,0){$\diamond$}}}
\put(418,375){\raisebox{-.8pt}{\makebox(0,0){$\diamond$}}}
\put(485,449){\raisebox{-.8pt}{\makebox(0,0){$\diamond$}}}
\put(659,368){\raisebox{-.8pt}{\makebox(0,0){$\diamond$}}}
\put(342,306){\raisebox{-.8pt}{\makebox(0,0){$\diamond$}}}
\put(422,260){\raisebox{-.8pt}{\makebox(0,0){$\diamond$}}}
\put(480,199){\raisebox{-.8pt}{\makebox(0,0){$\diamond$}}}
\put(501,153){\raisebox{-.8pt}{\makebox(0,0){$\diamond$}}}
\put(625,431){\raisebox{-.8pt}{\makebox(0,0){$\diamond$}}}
\put(1006,767){\makebox(0,0)[r]{'0.8'}}
\put(1050,767){\makebox(0,0){$\bullet$}}
\put(247,321){\makebox(0,0){$\bullet$}}
\put(267,328){\makebox(0,0){$\bullet$}}
\put(303,320){\makebox(0,0){$\bullet$}}
\put(520,527){\makebox(0,0){$\bullet$}}
\put(246,176){\makebox(0,0){$\bullet$}}
\put(710,340){\makebox(0,0){$\bullet$}}
\put(509,514){\makebox(0,0){$\bullet$}}
\put(264,149){\makebox(0,0){$\bullet$}}
\put(246,176){\makebox(0,0){$\bullet$}}
\put(710,340){\makebox(0,0){$\bullet$}}
\put(493,133){\makebox(0,0){$\bullet$}}
\put(461,127){\makebox(0,0){$\bullet$}}
\put(775,444){\makebox(0,0){$\bullet$}}
\put(356,295){\makebox(0,0){$\bullet$}}
\put(552,534){\makebox(0,0){$\bullet$}}
\put(238,216){\makebox(0,0){$\bullet$}}
\put(236,262){\makebox(0,0){$\bullet$}}
\put(239,300){\makebox(0,0){$\bullet$}}
\put(546,450){\makebox(0,0){$\bullet$}}
\put(348,127){\makebox(0,0){$\bullet$}}
\put(774,360){\makebox(0,0){$\bullet$}}
\put(595,404){\makebox(0,0){$\bullet$}}
\put(552,534){\makebox(0,0){$\bullet$}}
\put(238,216){\makebox(0,0){$\bullet$}}
\put(236,262){\makebox(0,0){$\bullet$}}
\put(595,404){\makebox(0,0){$\bullet$}}
\put(552,534){\makebox(0,0){$\bullet$}}
\put(766,484){\makebox(0,0){$\bullet$}}
\put(745,332){\makebox(0,0){$\bullet$}}
\put(503,146){\makebox(0,0){$\bullet$}}
\put(748,511){\makebox(0,0){$\bullet$}}
\put(765,339){\makebox(0,0){$\bullet$}}
\put(495,172){\makebox(0,0){$\bullet$}}
\put(715,526){\makebox(0,0){$\bullet$}}
\put(247,321){\makebox(0,0){$\bullet$}}
\put(267,328){\makebox(0,0){$\bullet$}}
\put(303,320){\makebox(0,0){$\bullet$}}
\put(356,295){\makebox(0,0){$\bullet$}}
\put(552,534){\makebox(0,0){$\bullet$}}
\put(238,216){\makebox(0,0){$\bullet$}}
\put(656,365){\makebox(0,0){$\bullet$}}
\put(520,527){\makebox(0,0){$\bullet$}}
\put(246,176){\makebox(0,0){$\bullet$}}
\put(238,216){\makebox(0,0){$\bullet$}}
\put(236,262){\makebox(0,0){$\bullet$}}
\put(239,300){\makebox(0,0){$\bullet$}}
\put(546,450){\makebox(0,0){$\bullet$}}
\put(604,535){\makebox(0,0){$\bullet$}}
\put(236,262){\makebox(0,0){$\bullet$}}
\put(239,300){\makebox(0,0){$\bullet$}}
\put(546,450){\makebox(0,0){$\bullet$}}
\put(604,535){\makebox(0,0){$\bullet$}}
\put(236,262){\makebox(0,0){$\bullet$}}
\put(595,404){\makebox(0,0){$\bullet$}}
\put(552,534){\makebox(0,0){$\bullet$}}
\put(766,484){\makebox(0,0){$\bullet$}}
\put(303,320){\makebox(0,0){$\bullet$}}
\put(356,295){\makebox(0,0){$\bullet$}}
\put(417,256){\makebox(0,0){$\bullet$}}
\put(467,210){\makebox(0,0){$\bullet$}}
\put(665,533){\makebox(0,0){$\bullet$}}
\sbox{\plotpoint}{\rule[-0.400pt]{0.800pt}{0.800pt}}%
\put(1006,722){\makebox(0,0)[r]{'0.6'}}
\put(1050,722){\raisebox{-.8pt}{\makebox(0,0){$\circ$}}}
\put(247,321){\raisebox{-.8pt}{\makebox(0,0){$\circ$}}}
\put(259,329){\raisebox{-.8pt}{\makebox(0,0){$\circ$}}}
\put(278,328){\raisebox{-.8pt}{\makebox(0,0){$\circ$}}}
\put(700,665){\raisebox{-.8pt}{\makebox(0,0){$\circ$}}}
\put(256,157){\raisebox{-.8pt}{\makebox(0,0){$\circ$}}}
\put(925,477){\raisebox{-.8pt}{\makebox(0,0){$\circ$}}}
\put(699,652){\raisebox{-.8pt}{\makebox(0,0){$\circ$}}}
\put(273,143){\raisebox{-.8pt}{\makebox(0,0){$\circ$}}}
\put(256,157){\raisebox{-.8pt}{\makebox(0,0){$\circ$}}}
\put(925,477){\raisebox{-.8pt}{\makebox(0,0){$\circ$}}}
\put(503,140){\raisebox{-.8pt}{\makebox(0,0){$\circ$}}}
\put(490,132){\raisebox{-.8pt}{\makebox(0,0){$\circ$}}}
\put(958,627){\raisebox{-.8pt}{\makebox(0,0){$\circ$}}}
\put(306,318){\raisebox{-.8pt}{\makebox(0,0){$\circ$}}}
\put(713,673){\raisebox{-.8pt}{\makebox(0,0){$\circ$}}}
\put(245,178){\raisebox{-.8pt}{\makebox(0,0){$\circ$}}}
\put(239,206){\raisebox{-.8pt}{\makebox(0,0){$\circ$}}}
\put(237,238){\raisebox{-.8pt}{\makebox(0,0){$\circ$}}}
\put(818,532){\raisebox{-.8pt}{\makebox(0,0){$\circ$}}}
\put(431,126){\raisebox{-.8pt}{\makebox(0,0){$\circ$}}}
\put(967,566){\raisebox{-.8pt}{\makebox(0,0){$\circ$}}}
\put(861,505){\raisebox{-.8pt}{\makebox(0,0){$\circ$}}}
\put(713,673){\raisebox{-.8pt}{\makebox(0,0){$\circ$}}}
\put(245,178){\raisebox{-.8pt}{\makebox(0,0){$\circ$}}}
\put(239,206){\raisebox{-.8pt}{\makebox(0,0){$\circ$}}}
\put(861,505){\raisebox{-.8pt}{\makebox(0,0){$\circ$}}}
\put(713,673){\raisebox{-.8pt}{\makebox(0,0){$\circ$}}}
\put(948,648){\raisebox{-.8pt}{\makebox(0,0){$\circ$}}}
\put(944,476){\raisebox{-.8pt}{\makebox(0,0){$\circ$}}}
\put(504,153){\raisebox{-.8pt}{\makebox(0,0){$\circ$}}}
\put(931,662){\raisebox{-.8pt}{\makebox(0,0){$\circ$}}}
\put(956,483){\raisebox{-.8pt}{\makebox(0,0){$\circ$}}}
\put(495,172){\raisebox{-.8pt}{\makebox(0,0){$\circ$}}}
\put(905,671){\raisebox{-.8pt}{\makebox(0,0){$\circ$}}}
\put(247,321){\raisebox{-.8pt}{\makebox(0,0){$\circ$}}}
\put(259,329){\raisebox{-.8pt}{\makebox(0,0){$\circ$}}}
\put(278,328){\raisebox{-.8pt}{\makebox(0,0){$\circ$}}}
\put(306,318){\raisebox{-.8pt}{\makebox(0,0){$\circ$}}}
\put(713,673){\raisebox{-.8pt}{\makebox(0,0){$\circ$}}}
\put(245,178){\raisebox{-.8pt}{\makebox(0,0){$\circ$}}}
\put(897,487){\raisebox{-.8pt}{\makebox(0,0){$\circ$}}}
\put(700,665){\raisebox{-.8pt}{\makebox(0,0){$\circ$}}}
\put(256,157){\raisebox{-.8pt}{\makebox(0,0){$\circ$}}}
\put(245,178){\raisebox{-.8pt}{\makebox(0,0){$\circ$}}}
\put(239,206){\raisebox{-.8pt}{\makebox(0,0){$\circ$}}}
\put(237,238){\raisebox{-.8pt}{\makebox(0,0){$\circ$}}}
\put(818,532){\raisebox{-.8pt}{\makebox(0,0){$\circ$}}}
\put(737,677){\raisebox{-.8pt}{\makebox(0,0){$\circ$}}}
\put(239,206){\raisebox{-.8pt}{\makebox(0,0){$\circ$}}}
\put(237,238){\raisebox{-.8pt}{\makebox(0,0){$\circ$}}}
\put(818,532){\raisebox{-.8pt}{\makebox(0,0){$\circ$}}}
\put(737,677){\raisebox{-.8pt}{\makebox(0,0){$\circ$}}}
\put(239,206){\raisebox{-.8pt}{\makebox(0,0){$\circ$}}}
\put(861,505){\raisebox{-.8pt}{\makebox(0,0){$\circ$}}}
\put(713,673){\raisebox{-.8pt}{\makebox(0,0){$\circ$}}}
\put(948,648){\raisebox{-.8pt}{\makebox(0,0){$\circ$}}}
\put(278,328){\raisebox{-.8pt}{\makebox(0,0){$\circ$}}}
\put(306,318){\raisebox{-.8pt}{\makebox(0,0){$\circ$}}}
\put(343,300){\raisebox{-.8pt}{\makebox(0,0){$\circ$}}}
\put(385,273){\raisebox{-.8pt}{\makebox(0,0){$\circ$}}}
\put(772,679){\raisebox{-.8pt}{\makebox(0,0){$\circ$}}}
\end{picture}